\documentstyle[aps,preprint,psfig]{revtex}
\begin{document}

\newcommand{\wa}{w_{0\rightarrow n}}
\newcommand{\wb}{w_{n\rightarrow 0}}
\newcommand{\wc}{w_{0\rightarrow w}}
\newcommand{\we}{w_{w\rightarrow 0}}

\title{Statistical Analysis of Native Contact Formation in the Folding of
Designed Model Proteins}
\author{Guido Tiana and Ricardo A. Broglia\\
        Department of Physics, University of Milano,\\
	via Celoria 16, 20133 Milano, Italy,\\
	INFN, Sez. di Milano, Milano, Italy,\\
	The Niels Bohr Institute, University of Copenhagen,\\
	Bledgamsvej 17, 2100 Copenhagen, Denmark}
\date{\today}
\maketitle

\begin{abstract}
The time evolution of the formation probability of 
native bonds has been studied
for designed sequences which fold fast into the native conformation. From this
analysis a clear hierarchy of bonds emerge a) local, fast forming highly stable
native bonds built by some of the most
strongly interacting amino acids of the protein, 
b) non--local bonds formed late
in the folding process, in coincidence with the folding nucleus, and involving
essentially the same strongly interacting amino acids already participating in
the fast bonds, 
c) the rest of the native bonds whose
behaviour is  subordinated, to a large extent, 
to that of the local-- and non--local
native contacts.
\end{abstract}

\section{Introduction}

There is strong experimental evidence which testifies to the fact that
the folding of small, single domain proteins can be pictured as the
crossing of a free energy barrier between the unfolded state U (or,
for some proteins like e.g. barnase \cite{fersht}, 
a populated intermediate state I which is fast at equilibrium 
with the unfolded state U) and the native state N \cite{fersht_book}. In
any case, experiments usually detect one single time scale, which can be
interpreted as
the characteristic time $\tau\sim\exp(\Delta F/kT)$ needed by the system
to cross the free energy barrier of height $\Delta F$ at a temperature
$T$, $k$ being the Botzmann constant.

Under such circumstances it would seem rather uninteresting to develop
models which describe the time evolution, in the folding process, of individual
native contacts, if nothing else because whatever results one would obtain
could be hardly checked experimentally, to the extent that the hierarchy of
events which could eventually be found should express themself through a
two--state picture.
On the other hand, while this barrier crossing picture describes 
correctly a great deal of small proteins, it is not helpful in explaining 
how proteins can find their native state on very short times (tipically ranging
from milliseconds to seconds) in spite of the enormous size of conformational
space. This problem is usually referred to as Levinthal's paradox
\cite{levinthal}.
To address this problem is the main goal of the present work.

We have found that designed proteins follow a well defined hierarchy of
steps {\it en route} to folding \cite{times}, which involve local elementary
structures and the formation of a folding nucleus \cite{sh_nucleus}.
These steps answer 
Levinthal's paradox providing at the same time a microscopic picture to the well
established two--state folding scenario. The final proof of the
correctness of our results is to be found in the fact that they allow us to
predict  the 3D--structure of any designed model protein from its amino
acid sequence \cite{1d3d}, provided the contact energies used to design the sequence
are known and an analytic expression which provide an overall account of the
formation probability of the elementary
structures involved in the folding process is available. The derivation 
of such an
analytic expression is also one the goals of this work.

In Section 2 we review the model used to design the notional protein. 
In Section 3 we
develop the analytic model of the time evolution of the native contacts and work
out the solutions of the associated master equations. The value of the
corresponding parameters are fixed by fitting these solutions with the numerical
results of the simulations. In Section 4 we extract,
from the results of the simulations,
an analytic expression for the formation probability of the bonds.
Chapter 5 contains the conclusions.

\section{The Model}

A useful theoretical approach to study protein folding is a simplified lattice
model, where the protein is a string of beads that is arranged on a cubic
lattice \cite{sh_nucleus,go,dill}. The configurational energy of a chain of N monomers is given by
\begin{equation}
E=\frac{1}{2}\sum^N_{i,j}U_{m(i),m(j)}\Delta(|\vec{r}_i-\vec{r}_j|),
\end{equation}
where $U_{m(i),m(j)}$ is the effective interaction potential between monomers
$m(i)$ and $m(j)$, $\vec{r}_i$ and $\vec{r}_j$ denote their lattice positions
and $\Delta(x)$ is the contact function. In Eq. (1) the pairwise interaction is
different from zero when $i$ and $j$ occupy nearest--neighbour sites, i.e.,
$\Delta(a)=1$ and $\Delta(na)=0$ for $n\geq 2$, where $a$ indicates the step
length of the lattice. In addition to these interactions, it is assumed that
on--site repulsive forces prevent two amino acids to occupy the same site
simultaneously, so that $\Delta(0)=\infty$ (excluded volume ansatz). 
The folding of the chain is simulated by Monte Carlo (MC) methods \cite{sklonik}.
We shall consider throughout a 20--letters
representation of protein sequence where $U$ is a $20\times 20$ matrix. A
possible realization of this matrix is given in ref. \cite{mj}
(Table VI), where it was derived from frequencies of contacts between different
amino acids in protein structures. The model we study here is a generic
heteropolymer model which has been shown to reproduce important generic features
of protein folding thermodynamics and kinetics, independent on the particular
potential chosen \cite{sh_opin,sh_fed}. This is achieved by using the same potential to design
sequences and to simulate folding. However, in using such an approach, one should
keep in mind that the labelling of amino acids (spherical beads all of the same
size and with no side chain) is generic too and may be no obvious relation
between those labels and labels for real amino acids.

Sequences are designed by minimizing, for fixed amino acid concentration, the
energy of the native conformation with respect to the amino acid sequence.
Good--folder sequences are characterized by a large gap $\delta=E_c-E_n$
(compared to the standard deviation $\sigma$ of the contact energies) between
the energy of the sequence in the native conformation $E_n$, and the lowest
energy of the conformations structurally dissimilar to the
native conformation \cite{sh_gap}. 

Examples of sequences designed to fold 
onto the conformation displayed in Fig. 1(a) are listed
in Table 1. Among them, the sequences displaying $E_n=-16.5$ in the units we are
considering ($0.6\; kcal/mol$) \cite{mj} has been studied extensively in the
literature \cite{sh_s36,klimov,my} and is denoted S$_{36}$. 
Very long, low temperature simulations
indicated that $E_c=-14$. Consequently, for the designed sequence S$_{36}$,
$\delta=2.5\gg\sigma=0.3$, a condition which guarantees fast folding \cite{sh_s36}. 
In fact, this sequence 
folds at $T=0.20$ in $8\cdot 10^8$ MC steps, while at $T=0.28$ it folds in the
even shorter time of $6.5\cdot 10^5$ MC steps. 
The fractional population of the native state
corresponding to these two temperatures is $91\%$ and $10\%$, respectively, to
be compared with a population of $0.5$ and  of $10^{-5}$ for the heteropolymer folding
temperatures of $T=0.25$ (temperature at which the probability for folding as well as for
unfolding is $1/2$) and $T=0.40$ (temperature at which bonds break essentially
as fast as they are formed due to thermal fluctuations) respectively. 
All the calculations
discussed below, unless explicitely mentioned,
 were carried out at the temperature $T=0.28$, optimal from the
point of view of allowing for the accumulation of representative samples of the
different simulations, and at the same time leading to a consistent population of
the native conformation.

We have discussed elsewhere \cite{times} the results of folding simulations of
designed sequences, which suggest the folding mechanism to be hierarchic. 
Let us here summarize the evolution of sequence S$_{36}$. 
The starting point is an elongated,
random generated conformation.  After $\sim 10^2$ MCS two things have
happened. The chain is now rather compact, having built of the order
of $20$ contacts (to be compared with the few contacts of the starting
denaturated
conformation and the maximum number of contacts for a full--compact
36mer, which is $40$), as a consequence of the negative contact energies of the
MJ interaction matrix (Table VI of ref. \cite{mj}). 
Also, three native contacts are already formed and rather stable (cf. Fig. 1(b)). They
are the contacts between monomers 3--6, 11--14 and 27--30, which are both
strongly interacting (their energy are $3.4$ kT, $2.7$ kT and $3.4$ kT,
respectively) and local (that is, close along the chain).

Afterwards, the system searches among rather compact conformations, 
until the native
contacts among the local elementary structures (LES) 
formed by monomers 3--4--5--6, 27--28--29--30 and 11--12--13--14 
 (which we  call S$_4^1$, S$_4^2$
and S$_4^3$, respectively \cite{aggreg}) 
are built. This is the post--critical "folding nucleus"
\cite{sh_nucleus} 
and its formation takes $\sim 7\cdot 10^5$ MCS. Once this is
formed, the chain can reach the native conformation in $\sim 1000$ MCS.
The key element determining the overall folding time, is the
formation of the (post--critical) folding nucleus. 
However, this does not happen in an all--none fashion, but in a much more subtle
and elaborated way, as testified by the dynamics of native bonds formation shown
in Fig. 1(b). In fact, in the particular (and representative) case of S$_{36}$,
the folding nucleus is built in two steps, after the LES have been formed.
First, the two LES S$_4^1$ and S$_4^2$ come together and occupy their native
relative conformation in $\approx 10^5$ MCS.
In the successive $4-5$ hundred thousands steps the partially formed folding
nucleus is surrounded by non--native neighbours. When the remaining LES 
S$_4^3$ finds its native position, the folding nucleus is completed
and the chain folds almost immediately afterwards.

\section{Statistical Analysis of Native Contact Formation}

To provide a quantitative basis to discuss the behaviour observed 
in the snapshots
of folding events discussed in the previous section, we have carried out
a systematic study of native bonding statistics.
For this purpose, the main quantity studied 
was the probability $P_{i,j}(t)$ that
the $i$th monomer forms its native contact with the $j$th monomer
as function of time, quantity found as average over many time evolutions. In
every time evolution, the chain starts from a random conformation
and the simulation is carried out over $10^6$ MC steps.

From this analysis we have found that the $40$ native bonds of sequence S$_{36}$  
can be classified in three groups according to the behaviour of the
associated function $P_{i,j}(t)$ (see Fig. 2(b), (c) and (d) ).
The first group is composed of bonds which
reach a stability considerably higher than all the other
contacts ($P_{i-j}\approx 1$)
on short call. Bonds $3-6$, $11-14$ and $27-30$
(which we call "fast bonds") belong to this class and become stable in a
time scale of the order of $10^3$ MCS.
A second set of bonds (which we call "slow bonds") acquire a $0.5-0.8$
stability  in
a time scale of order of $10^4-10^5$ MC steps.
The majority of bonds reach
a stability smaller than $0.5$ in a time scale which is again of the order
of $10^5$ MC steps.

In order to understand the behaviour of $P_{i,j}$, we assume that each
bond $i-j$ can be found in one of three possible states. It can be in its
native state (which we will denote "N"), it can be unbound ("0") or
its residues can be bound to some non--native residues ("W"). We also
assume that the system can move between "0" and "N", between "0" and
"W", but not between "N" and "W". The state "0" will denote not only
the cases in which monomers $i$ and $j$ have no nearest neighbours
at all, but also the cases in which their nearest neighbours interact
weakly. This is the case when in a globule conformation a monomer
moves to break a bond. If the density of monomers is high, in fact,
it is unlikely that it can move into an isolated site. Nonetheless, this
monomer will become neighbour of monomers which have no energetic
preference for it (so they can be considered "random" monomers) and
consequently the interaction energy between the moved monomer and its new
neighbours is, in first approximation, zero. 

According to this model it is possible
to write, for each bond, the master equations
\begin{eqnarray}
\dot{P_n}(t)&=&\wa P_0(t) - \wb P_n(t), \\ \nonumber
\dot{P_w}(t)&=&\wc P_0(t) - \we P_w(t), \\
P_0(t)&=&1-P_n(t)-P_w(t), \nonumber
\end{eqnarray}
where $P_n(t)$, $P_w(t)$ and $P_0(t)$ are the probabilities that the bond
under study is in its native state, in a non--native state or in a
unbound state, respectively. The initial conditions for folding are that
bonds start in an unbound state, so that $P_0(0)=1$.

To solve Eqs. (2), it is necessary to know how the transition
probabilities $\wa$, $\wb$, $\wc$ and $\we$ for a given bond
depend on the states of all other bonds. We will show that the
transition probabilities associated with "fast" bonds 
are essentially independent on the state of
the other bonds, while the probabilities associated with "slow" bonds
depend on the state of "fast" bonds. However, since "fast" bonds reach an
equilibrium probability close to $1$ in a time scale which is three
orders of magnitude faster than that associated with 
"slow" bonds, it is possible to
neglect this dependence. Although the situation is more complicated for
the other bonds, as it will be clear later, the folding rate
of the designed protein is essencially independent of them. In any case, one should
notice that the transition probabilities $\wa$, $\wb$, $\wc$ and $\we$ represent
average values, and consequently do not depend on the actual conformation.

The probability that a native bond is formed is given by the solution of
Eqs. (2), that is,
\begin{equation}
P_n(t)=\alpha (1-\exp(-\lambda t))+\beta(1-\exp(-\mu t)),
\end{equation}
where 
\begin{eqnarray}
\lambda&=&\frac{1}{2}\left(\wa+\wb+\wc+\we+\right.\\\nonumber
 &&+\left.\left((\wa+\wb+\wc+\we)^2-\right.\right.\\\nonumber
 &&+\left.\left.4(\wa\we+\wb\we+\wb\wc)\right)^{1/2}\right),
\nonumber
\end{eqnarray}
\begin{eqnarray}
\mu&=&\frac{1}{2}\left(\wa+\wb+\wc+\we-\right.\\\nonumber
&&+\left.\left((\wa+\wb+\wc+\we)^2\right.\right.-\\\nonumber
&&+\left.\left.4(\wa\we+\wb\we+\wb\wc)\right)^{1/2}\right),
\nonumber
\end{eqnarray}
\begin{equation}
\alpha=\frac{1}{\lambda-\mu}\left(\frac{-\mu\wa\we}{\wa\we+\wb\we+\wb\wc}+\wa\right),
\end{equation}
\begin{equation}
\beta=\frac{1}{\lambda-\mu}\left(\frac{\lambda\wa\we}{\wa\we+\wb\we+\wb\wc}-\wa\right).
\end{equation}
The set of parameters $\lambda$, $\mu$, $\alpha$ and $\beta$ associated with
each of the native bonds of sequence S$_{36}$ were determined through a
least--square fit of the results of the MC simulations (cf. e.g. Figs. 2(b),
2(c) and 2(d) ). From these values of $\lambda$, $\mu$, $\alpha$ and $\beta$ the
transition probabilities $\wa$, $\wb$, $\wc$ and $\we$ were determined with the
help of Eqs. (4--7).

\subsection{Fast Bonds}

The bonds between monomers $3-6$, $11-14$ and $27-30$ are formed remarkably
fast. The parameters of Eq. (2), fitted in the range
$0-1000$ of MC steps, and the transition
rates derived from them are shown in Table 2.
The standard deviation between the fitting function and the data is
$9.2\cdot 10^{-5}$, indicating that Eq. (2) describes accurately
the bonding of these monomers.
According to this relation the bonding process takes place with two
different characteristic times, namely $1/\lambda$ and $1/\mu$,
depending on whether the monomers involved can form the bond directly, or
get trapped in non-native bonds ("w"). The order of magnitude of the
overall process can be summarized by
the characteristic time $\tau=(\alpha\lambda)^{-1}+(\beta\mu)^{-1}$.
The characteristic times $\tau$ for fast bonds under discussion 
are of the
order of $10^2$ MCS (Table 3), a very short
time as compared to the average first passage 
folding time of $7\cdot 10^5$ MC steps.

The effective bonding free energy $\Delta
F$ can be derived from the expression $\wb=w_0 \exp(\Delta F/T)$,
$w_0$ being the inverse of the time step. Since at each MC step before
folding,
every monomer is attempted to move with equal probability\cite{footnote1},
it is reasonable to chose $w_0=2/N$, $N$
being the length of the chain and the factor $2$ keeps into account that
it is possible to create or break a bond moving one of its two monomers.
The values of $\Delta F$ for the three "fast" bonds are $\Delta F=-1.05$,
$\Delta F=-1.01$ and $\Delta F=-1.15$, respectively.
The bonding free energy is $\Delta F=F_n-F_0=E_n-E_0-TS_n+TS_0$, where
$E_n$, $E_0$ ($=0$), $S_n$ ($=0$) and $S_0$ are the contact 
energies and entropies associated with the native and
with the unbound states. In the case of bond 3--6, where 
$E_n-E_0=-0.97$ (similar values
are associated with the other fast bonds), one obtains $S_0\approx 0$.
Consequently, there is in average only one non--bound state available
to the system, due to the excluded volume constrain (to be noted that
in the present simulations, the chain starts from a random generated
conformation, which is usually rather compact). If the chain was in a 
non--compact conformation, there would be 9 
possible unbound states for bonds between monomers $i$ and $i+3$, 
which would lead to an entropy equal to
$S_0=2.19$ and, consequently, to an overall stability of
$\alpha+\beta=0.35$ (instead of $0.77$). According to the present model,
it is necessary a collapse of the chain to a compact conformation before the
fastest native bonds are stabilized
\cite{footnote2}

On a time scale of $10^5$ MC steps, fast bonds 3--6, 11--14 and 27--30
reach a stability
of $\alpha+\beta=0.97$, $\alpha+\beta=0.91$ and $\alpha+\beta=0.96$,
respectively (instead of $0.77$, $0.76$ and $0.65$, as measured on a $10^3$
timescale). This increment
in stability can only come from the interaction with other monomers
which is, as we shall see, the case.

\subsection{Slow Bonds}

The distributions $P_n(t)$ related to the bonds acting among the set of monomers
$(3,4,5,6)$, $(11,12,13,14)$ and $(27,28,29,30)$
(which we have called "slow" bonds)
display a remarkable similarity. The parameters associated with some of these
distributions are listed in Table 4. Again, the functions $P_n$ are well described
by Eq. (2), the average standard deviation from the fitted function being
$2.2\cdot 10^{-5}$. The dependence of the transition rates associated with
these bonds on the state of "fast" bonds is negligible, since "fast" bonds
reach equilibrium (with $P_w(\infty)\sim 1$) in a time which is 3 orders
of magnitude shorter than the relaxation time of "slow" bonds. The
characteristic bonding times $\tau$ for these bonds range from $4.4\cdot
10^4$ to $2.8\cdot 10^5$
MC steps.

The unbinding free energy for the "slow" bonds, calculated in the same way
as for the "fast" bonds, ranges from $-1.57$ to $-2.43$. The bond $28-5$, for
example, has $\Delta F=-2.61$ and the monomers interact with a potential
energy $E=-0.41$. Assuming again that the native state is unique, this bond
should satisfy
\begin{equation}
-0.41+TS_0=-2.61
\end{equation}
which is impossible, because it gives $S_0=-7.3$ (and $S_w=-1.93$, setting
$E_w\approx-\sigma=-0.3$) 
while the entropy of the unbound state $S_0$ must be  a positive quantity.
On the other hand, we can consider that the monomers $(3,4,5,6)$ form a rigid
structure (the bond $3-6$ is very stable) that interact, as a whole, with 
the structure built of the monomers $(27,28,29,30)$ (the bond $27-30$ is also
very stable). In this case, the total energy involved is the sum of the four
energies of the bonds associated with the 8 monomers contained in
the two structures, that is $E=-3.02$. Thus, Eq. (8) becomes
\begin{equation}
-3.02+TS_0=-2.61,
\end{equation}
which is satisfied if $S_0=1.46$, leading to $\approx 4$ unbound states. 
The same argument applies when considering
the interaction, as a whole,  between the monomers $(11,12,13,14)$ and those belonging to the
other two structures. 

From the above results, one identify the "fast" bonds with 
those which stabilize the local elementary
structures discussed in Section 2 
and "slow" bonds as those among local elementary
structures.

\subsection{The other bonds}

So far, only very favorably interacting bonds have been studied, whose
interaction energy is between $-0.7$ and $-1$. The majority of bonds,
however, have higher energy, sometimes even positive. In Fig. 2(d) it
is displayed the behaviour of $P(t)$ associated with 
the bond 24--15, whose interaction
energy is $-0.38$, as a function of time. 
The fit with the function given in Eq. (2) leads to
$\wb=1.3\cdot 10^{-5}$. It follows that the free energy which keeps
the two residues bound is $\Delta F=-2.14$, quantity which can be decomposed
as $\Delta F=E_n+TS_0$, in keeping with the fact 
that $E_0=S_n=0$. This cannot hold for any
positive value of $S_0$, implying that there is an effective force, due to the
correlation with other residues, which
decreases the value of $E_n$ below $-0.38$. 

Since this class of weakly interacting bonds, unlike "slow bonds",
are not associated with stable substructures, correlations among them
can only be due to the polymeric constrain,  after "fast" and
"slow" bonds have reached the native state, building a stable
folding nucleus.
Under such circumstances the remaining 
monomers form short loops which begin and end in the folding nucleus.
The space available
to these residues is small and the polymeric bonds keep them together
in such a way that each loop interact cooperatively with the rest of
the chain (even if they are less rigid than the local
elementary structures).
In the case of bond 24--15, for example,
monomer 15 is almost fixed once the core has been stabilized, being
constrained, to a large extent, 
by residue 14 belonging to the core. Monomer 24
belongs to a loop which extends from monomer 16 to monomer 27, and
is thus affected by the interactions of all these monomers
among themselves and with the core.

If this scenario is correct, then the fit of $P(t)$ obtained making use of 
Eqs. (3--7) is
not correct, since the parameters of the master equations
now depend on time, leading to a more complicated solution. Assuming
for example that the rate $\wa$ has a time dependence in the form
$\wa(t)=\wa'(1-\exp(-\nu t))$, with $\nu=4\cdot 10^{-5}$, the typical
formation time of slow bonds, we found $\Delta F=-1.58$.
The characteristic bonding time for the bond 24--15 is $\tau=2.6\cdot10^5$
(other times are, i.e., $\tau=3.3\cdot 10^5$ for bond 7--10 and
$\tau=3.2\cdot 10^5$ for bond 10--15). This is very fast, considering
that in that time it is included the bonding time of "slow" bonds,
on which it depends. At the basis of this result one finds the 
decrease in entropy caused
by the formation of the folding nucleus, an event which
not only contribute to stabilizing these
weak bonds, but also make their formation faster.

Summing up, the chain first builds strongly interacting local contacts
(3--6, 11--14 and 27--30) in a time which is of the order of 100 MCS.
The structures built in this way ("local elementary structures") 
behave as elementary entities,
which are strongly interacting, have a small conformational space available
and are unlikely to form stable non--native bonds. These local
elementary structures come
to their native position in a time of the order of $10^5$ MCS, building
the folding nucleus of the protein. Almost immediately 
hereafter, the weak residues
fold to their native position, due to the reduction of the conformational
space available to them and to the strong effective force that the
polymeric chain produces.

\subsection{Hindsight}

Folding seems to take place as a hierarchical succession of events.
Why this succession of events makes folding fast? Let us consider the number of
conformations available at each level of the folding process (cf. Fig. 3). 
At the beginning there 
are no constrains to the system and the chain can assume $10^{24}$ different
conformations. After the chain has collapsed to a compact
globule, there are $10^{19}$ conformations. The formation of the LES,
which is a local event, and consequently does not require 
 a massive search in conformation
space, further reduces the number of conformations to $10^{14}$. The native 
bonding of
LES  S$_4^1$ and S$_4^2$ further decrease the size of
conformational space to $10^{11}$, while only $10^5$ conformations are
associated to the folding nucleus.
Consequently, at each step of the hierarchy of events bringing the heteropolymer
from the denaturated state to the native state, the
system has to search only among a number of conformations given by the ratio between
the number of conformations associated with the step in question and the
successive step. In
this way  entropy is slowly squeezed out of the chain, until the system
reaches the unique
ground state, with zero entropy. One
can notice that the most important step in terms of the decrease in 
entropy of the system is the step associated with the completion of the
folding nucleus, i.e. the step in which the LES S$_4^3$ forms its native
contacts with the partially formed nucleus made out of the LES 
S$_4^1$ and S$_4^2$,
during which the system has to search among $10^6$
conformations. To be noted that to this  step  
is associated a rather small gain in energy 
(2.13, to be compared with 3.02 associated with the binding of local
elementary structures
 S$_4^1$ and S$_4^2$). This is likely the top of the free energy barrier separating the native from the unfolded states.

The importance in the folding process of a designed sequence of local
elementary structures is multiple, in that 
they decrease the entropy of the chain, enhance the
elementary contact energies (from pairs of monomers to pairs of LES) and
reduce the number of possibilities in which the most strongly
interacting residues can assemble together. 
The decrease in entropy is due to the fact that elementary structures are
rather stable, so that all the residues belonging to an elementary substructure
behave effectively as a single residue and the chain becomes, effectively,
shorter.
Moreover, while contact energies among the single, strongest interacting, amino
acids are of the order of $-1=-3.6 kT$ \cite{mj}, the interaction between LES built
out of four of these amino acids (e.g. S$_4^1$ and S$_4^2$) is four times
stronger ($\approx-4=-13 kT$). Finally, wrong (non--native) contacts are more
probable to be established during the folding process between single strongly
interacting monomers, than between local elementary structures.
 The set of local elementary structures is like 
a 4--digits code. To create a strong bond, all the
four monomers have to match.
After the local elementary structures have come to occupy
their native position in the protein, the chain is composed
by a nucleus surrounded by short loops. Each loop
can build its native bonds in a very short time due to its small entropy, leading the chain to the native
state.

For other sequences,
the same mechanism has been found to be at the basis of the folding process, 
even if the LES can now assume
different shapes. For example, folding of sequence number 8 of Table 1, 
to the
conformation shown in Fig. 1(a), is controlled by the local elementary 
structures formed by monomers 1--2, 16--21 and
31--32. In this case, a LES is larger than those associated with S$_{36}$, involving 6
monomers, and two of them are "degenerate", being composed of just two monomers, the
peptide bond playing the role of the strong interaction which gives life to the
elementary substructure. The case of the sequence folding to the conformation displayed in
Fig. 4(a) (cf. caption to the figure) is rather peculiar, in the sense that this native conformation has been
design to minimize the amount of local contacts \cite{sh_local}. Also in this
case folding is lead by local elementary structures, which are 1--6, 30--31 and 20--22 (cf.
Fig. 4(b) ). A further example, associated with a chain made of 48
monomers, is displayed in Fig. 4(c). To the sequence folding to this
conformation and listed in the caption, correspond rather large
local elementary structures,
built of monomers 2--12 and 33--41 (in Fig. 4(d) it is displayed the bond
dynamics for this sequence).

\section{Discussion}

The results collected in 
Tables 2 and 4 testify to the fact that the two main features which
characterize the contacts as belonging to the three different groups 
of native bonds are their
locality (namely, their distance along the chain) and their interaction
energies. "Fast bonds" are local and strongly interacting, "slow bonds" are
non--local and strongly interacting, while the other bonds are weakly
interacting.

Whether a bond belongs to a group or to another is determined by the values
of the jumping rates $\wa$, $\wb$, $\wc$ and $\we$.
The rate $\wa$ can be thought as having a purely geometrical
dependence. In Fig. 5 we display the inverse of $\wa$ associated with
the sequence S$_{36}$ as a
function of the distance $|j-i|$, measured along the chain, between the two
monomers involved in the bond. The monotonic behaviour of the values of $\wa$
 is compared with the inverse probability that the $i$th and $j$th
monomers come close during a random search (dotted curve in Fig. 5).
Such probability is simply
the number of conformations $\gamma^{N-2} (j-i)^{-1.68}$, $\gamma$ being  the
effective coordination number of the cubic lattice \cite{flory},
in which the $i$th monomer is nearest neigbours
of the $j$th monomer 
divided by the total number of conformations $\gamma^{N-1}$ of the chain.
The above expression for the number of conformations with a bond between the
$i$th and $j$th monomer is given by the product of three terms associated with
the corresponding parts which compose each conformation, namely two branches of lenght
$i$ and $N-j+1$ respectively, and a loop of length $j-i$. The first two terms 
contribute a factor
$\gamma^{i-1}$ and $\gamma^{N-j}$ respectively. The contribution
$\gamma^{j-i-1}(j-i)^{-1.68}$
associated with the loop was determined through a fit of the number of conformations
obtained by a complete enumeration up to\cite{footnote3} 
$j-i=16$ \cite{anke}.

While it is not easy to derive $\wc$ from theoretical considerations,
it is  correlated to $\wa$ (at least, as order of magnitude (see Tables 2 
and 4)). On the other hand, $\we$ is approximately constant ($1.9\cdot
10^{-2}$ for fast bonds and $4.6\cdot 10^{-5}$ for the
other bonds).

The parameter $\wb$ depend on the free energy difference between
the native bond and the unbound state. For "fast" bonds the change
in internal energy is given by the associated matrix elements,
depending only on the kind of the  pairs of monomers involved. Apparently,
the change in entropy between unbound and native states
is negligible for these bonds. In the case
of "slow" bonds, that is bonds among local elementary structures, 
one has to consider
the change in internal energy of all the monomers belonging to the
structures, while the entropy is rather constant ($S_0=-2.43\pm 0.09$).

\section{Conclusions}

The time evolution of the probability of native bond formation has been studied
for designed sequences which fold fast into the native conformation. From this
analysis we have found that the native bonds can be classified in three groups
namely, ("fast") bonds which reach a stability considerably higher than all the
other contacts (and close to 1, at "biological" temperatures) in a time scale of
the order of $10^3$ MC steps, ("slow") bonds which acquire $0.5-0.8$ stability
in a time scale of the order of $10^4-10^5$ MC steps, and bonds (the majority)
which reach a stability smaller than $0.5$ in a time scale which is again of the
order of $10^5$ MC steps. The first two types of bonds correspond to the (few)
local and the (relatively many) non--local contacts found in the (post--critical) folding nucleus.
Fast and slow contacts completely control the folding process. These results
provide a simple picture of the folding process, where local elementary
structures, stabilized by the strong, fast native bonds build, by assembling
together, the slow bonds (folding nucleus), after which the remaining bonds fall
in place within a very short time.

\bigskip\bigskip
{\Large Acknowledgments}\bigskip

We are grateful to Anke Odermann for the data concerning the full enumeration used in
Chapt. 4.

\newpage

\begin{table}
\begin{tabular}{|l|l|l|}\hline
1 & -17.13 & YPDLTKWHAMEAGKIRFSVPDACLNGEGIRQVTLSN \\
2 & -16.50 & SQKWLERGATRIADGDLPVNGTYFSCKIMENVHPLA \\
3 & -16.38 & GNRLPESGAAKGHELDFAWGTLVLSQKYIDNMRVPS \\
4 & -16.32 & TGKVQEGAIERSNDLDMAAGTLCIHPKPLESWRVYN \\
5 & -15.27 & NQEPLKRNGRDCARGTLYVSHGPFDVEIMKTIAWLA \\
6 & -15.24 & PPKVLERQTGNAINGDFDAYRGSCWLKLVEHTSAIM \\
7 & -14.91 & PENLFERQVWHTGDMDIPCYRVGLSAKGINKLTASA \\
8& -14.06 & DKYAIPDRTTNLVNGEFHVKRGCGSMQLSPELWAIA \\
\hline

\end{tabular}
\caption{Eigth sequences obtained through a Metropolis Monte Carlo
optimization in the space of sequences at different "selective"
temperatures. The conformation onto which they have been minimized
is the one displayed in Fig. 1(a). The second sequence of the list
is known in the literature [14-16] as S$_{36}$. }
\end{table}

\begin{table}

\begin{tabular}{|c|c|c|c|c|}
\hline
bond & $\alpha$ & $\lambda$ & $\beta$ & $\mu$ 
\\ \hline
3-6 & 0.19 & 0.15 & 0.58 & 0.0047 \\
27-30 & 0.25 & 0.07 & 0.51 & 0.0049  \\
11-14 & 0.38 & 0.023 & 0.27 & 0.0017 \\\hline
\end{tabular}

\begin{tabular}{|c|c|c|c|c|}
\hline
bond & $\wa$ & $\wb$ & $\wc$ & $\we$
\\ \hline
3-6 & $3.1\cdot10^{-2}$ & $1.3\cdot10^{-3}$ &
$1.0\cdot10^{-1}$ &  $1.7\cdot10^{-2}$\\
27-30 & $1.9\cdot10^{-2}$ & $1.5\cdot10^{-3}$ &
 $4.0\cdot10^{-1}$ & $1.3\cdot10^{-2}$ \\
11-14 & $9.1\cdot10^{-3}$ & $9.3\cdot10^{-4}$ &
 $1.1\cdot10^{-2}$ & $2.7\cdot10^{-3}$ \\\hline
\end{tabular}

\caption{The fitted parameters and the related jumping rates for "fast"
bonds done on the time range $0-1000$ MC steps,
derived from these parameters making use of Eqs. 4--7. The standard
deviation of the fitted curve from the data is $9.2\cdot 10^{-5}$.}
\end{table}

\begin{table}
\begin{tabular}{|c|c|c|c|}
\hline
bond  & $\tau$ & bond & $\tau$ \\\hline
3-6 & $401$ & 14-27 & $1.7\cdot 10^5$ \\
27-30 & $457$ & 6-11 & $2.5\cdot 10^5$ \\
11-14 & $2292$ & 28-13 & $1.9\cdot 10^5$ \\
28-5 & $1.3\cdot 10^5$ & 15-24 & $2.6\cdot 10^5$ \\
27-6 & $1.2\cdot 10^5$ & 7-10 & $3.3\cdot 10^5$\\
4-29 & $4.4\cdot 10^4$ & 10-15 & $3.3\cdot 10^5$ \\
5-12 & $2.0\cdot 10^5$ & &\\\hline
\end{tabular}

\caption{Characteristic times for some selected bonds.}
\end{table}
\begin{table}

\begin{tabular}{|c|c|c|c|c|}
\hline
bond & $\alpha$ & $\lambda$ & $\beta$ & $\mu$ \\\hline
28-5 & 0.17 & $1.3\cdot 10^{-4}$ & 0.61 & $1.9\cdot 10^{-5}$ \\
27-6 & 0.21 & $1.4\cdot 10^{-4}$ & 0.60 & $2.0\cdot 10^{-5}$ \\
4-29 & 0.44 & $7.3\cdot 10^{-5}$ & 0.39 & $1.9\cdot 10^{-4}$  \\
5-12 & 0.26 & $9.7\cdot 10^{-4}$ & 0.24 & $2.1\cdot 10^{-5}$ \\
14-27 & 0.25 & $1.4\cdot 10^{-4}$ & 0.40 & $1.7\cdot 10^{-5}$ \\
6-11 & 0.46 & $7.8\cdot 10^{-4}$ & 0.21 & $1.9\cdot 10^{-5}$ \\
28-13 & 0.11 & $3.7\cdot 10^{-4}$ & 0.34 & $1.7\cdot 10^{-5}$ \\\hline
\end{tabular}

\end{table}
\begin{table}

\begin{tabular}{|c|c|c|c|c|}
\hline
bond & $\wa$ & $\wb$ & $\wc$ &
$\we$ \\\hline
28-5 & $3.3\cdot
10^{-5}$ & $4.9\cdot 10^{-6}$ & $5.3\cdot 10^{-5}$ & $5.7\cdot 10^{-5}$ \\
27-6 &  $4.1\cdot
10^{-5}$ & $4.6\cdot 10^{-6}$ & $5.9\cdot 10^{-5}$ & $5.4\cdot 10^{-5}$ \\
4-29 & $3.1\cdot
10^{-5}$ & $1.0\cdot 10^{-5}$ & $1.0\cdot 10^{-5}$ & $1.6\cdot 10^{-5}$ \\
5-12  & $2.5\cdot
10^{-4}$ & $1.4\cdot 10^{-5}$ & $6.8\cdot 10^{-4}$ & $3.9\cdot 10^{-5}$ \\
14-27 & $4.8\cdot
10^{-5}$ & $7.7\cdot 10^{-6}$ & $7.0\cdot 10^{-5}$ & $3.7\cdot 10^{-5}$ \\
6-11  & $3.6\cdot
10^{-4}$ & $1.1\cdot 10^{-5}$ & $3.9\cdot 10^{-4}$ & $2.7\cdot 10^{-5}$ \\
28-13 & $4.6\cdot
10^{-5}$ & $1.0\cdot 10^{-5}$ & $2.6\cdot 10^{-4}$ & $6.0\cdot 10^{-5}$ \\\hline
\end{tabular}

\caption{The parameters related to the "slow" bonds, fitted in the range
$0-10^6$ MC steps. The typical standard deviation for these fits is
$2.2\cdot 10^{-5}$.}
\end{table}

\newpage

\begin{figure}
\centerline{\psfig{file=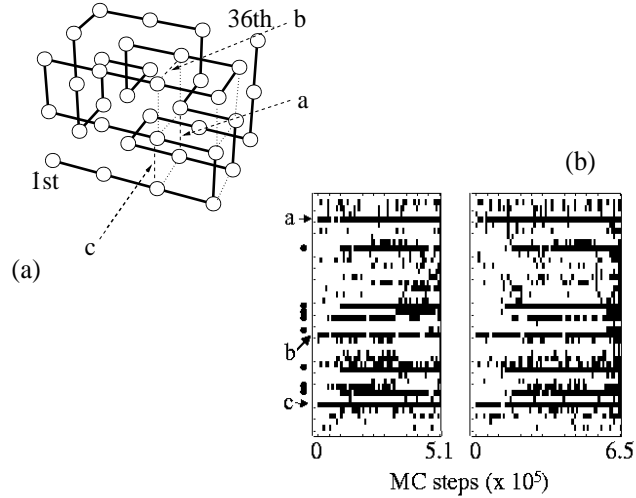,height=7cm,width=8.6cm}}
\caption{(a) The native conformation of sequence
SQKWLERGATRIADGDLPVNGTYFSCKIMENVHPLA (upper left). Bonds within local
elementary
structures (LES) are indicated with a dashed line, while bonds
among local substructures with a dotted line. Examples of bonds among LES
involve monomers
27--30, 11--14 and 3--6, and are marked a, b and c, respectively.(b) The
dynamics of native contacts for two typical runs 
is shown in the lower right panels. On the x-axis is time, while to the 
y-axis are associated the 40 native contacts that this sequence can build.
A black square in the plot indicates that a given bond is built at that
time. Bonds stabilizing local elementary structures are marked 
with a, b and c,
while bonds among local elementary structures 
are marked with solid dots. }
\end{figure}

\begin{figure}
\centerline{\psfig{file=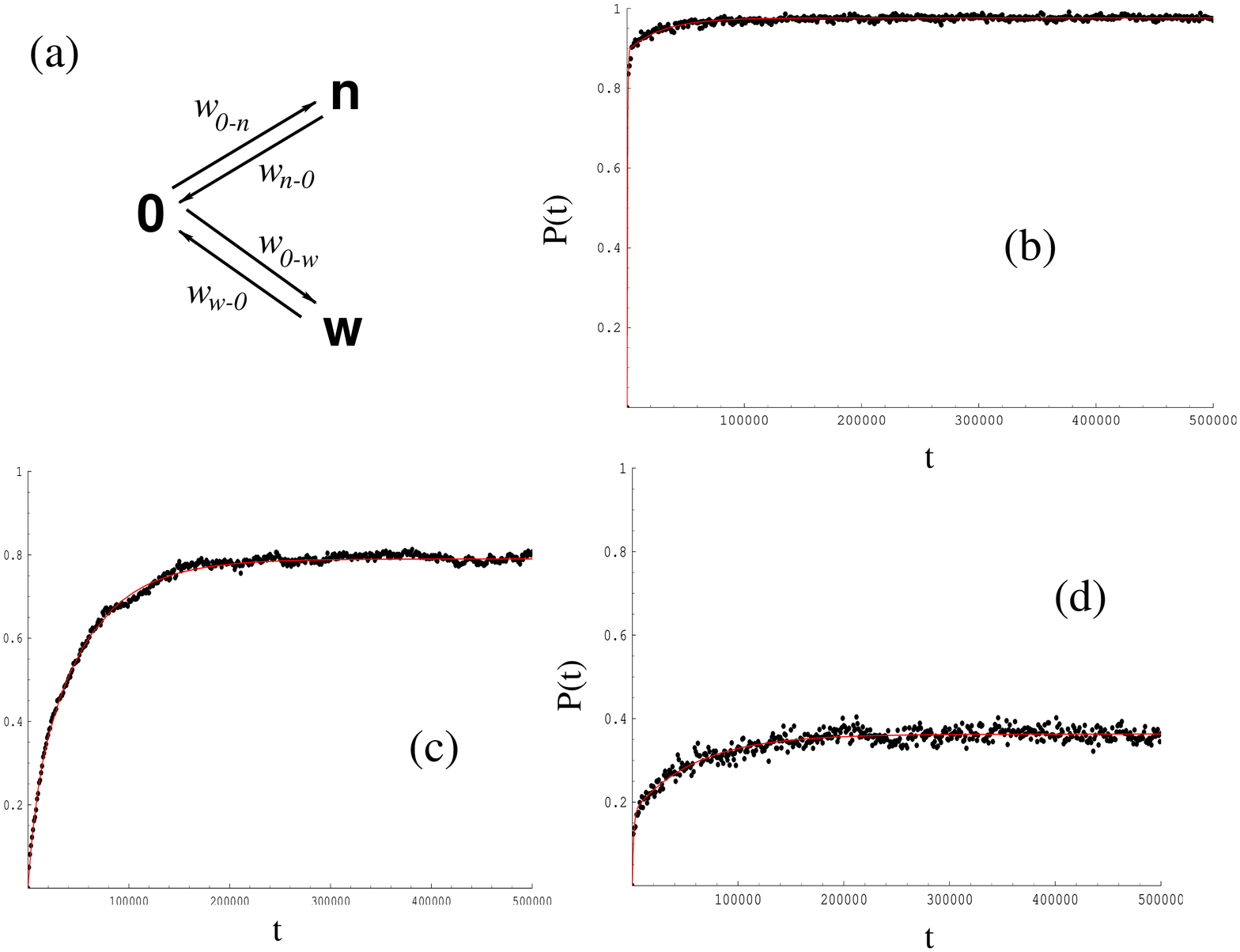,height=7cm,width=8.6cm}}
\caption{(a) Each native contact can be though to be in its native
state ("n"), in an unbound state ("0") or bound to some non--native
monomer ("w"). This model is described by Eq. (1). (b) The probability
$P_{i-j}(t)$ for the bond 3--6. To the curve obtained by simulations
is superimposed the least--squares fit done with Eq. 2.
(c) the same for bond 5--28. (d) The same for bond 15--24.}
\end{figure}

\begin{figure}
\centerline{\psfig{file=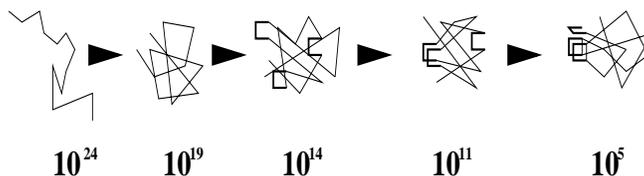,height=2.2cm,width=8.6cm}}
\caption{Schematic representation of the mechanism according to which the system sqeezes out its entropy, by
reducing the number of conformations available to it
(indicated in lower part of the
figure). The pieces of the heteropolymer drawn with heavy lines indicate the LES
of the model protein.}
\end{figure}

\begin{figure}
\centerline{\psfig{file=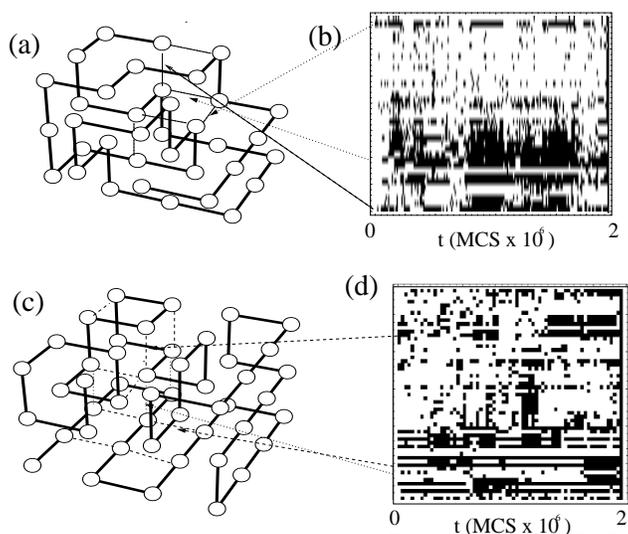,height=7cm,width=8.6cm}}
\caption{The native states of the sequence
RASM\-KDKTV\-GIGHQ\-LYLNFE\-GEWCPA\-PDNTRV\-SLAI (a) and
IMESQKWLCMEPAHWCVYTIQGLGNVNCPNTREFDSGRSKIQDAYLFH (c). The dynamics of the native bonds for
these sequences are displayed in (b) and (d), respectively.}
\end{figure}

\begin{figure}
\centerline{\psfig{file=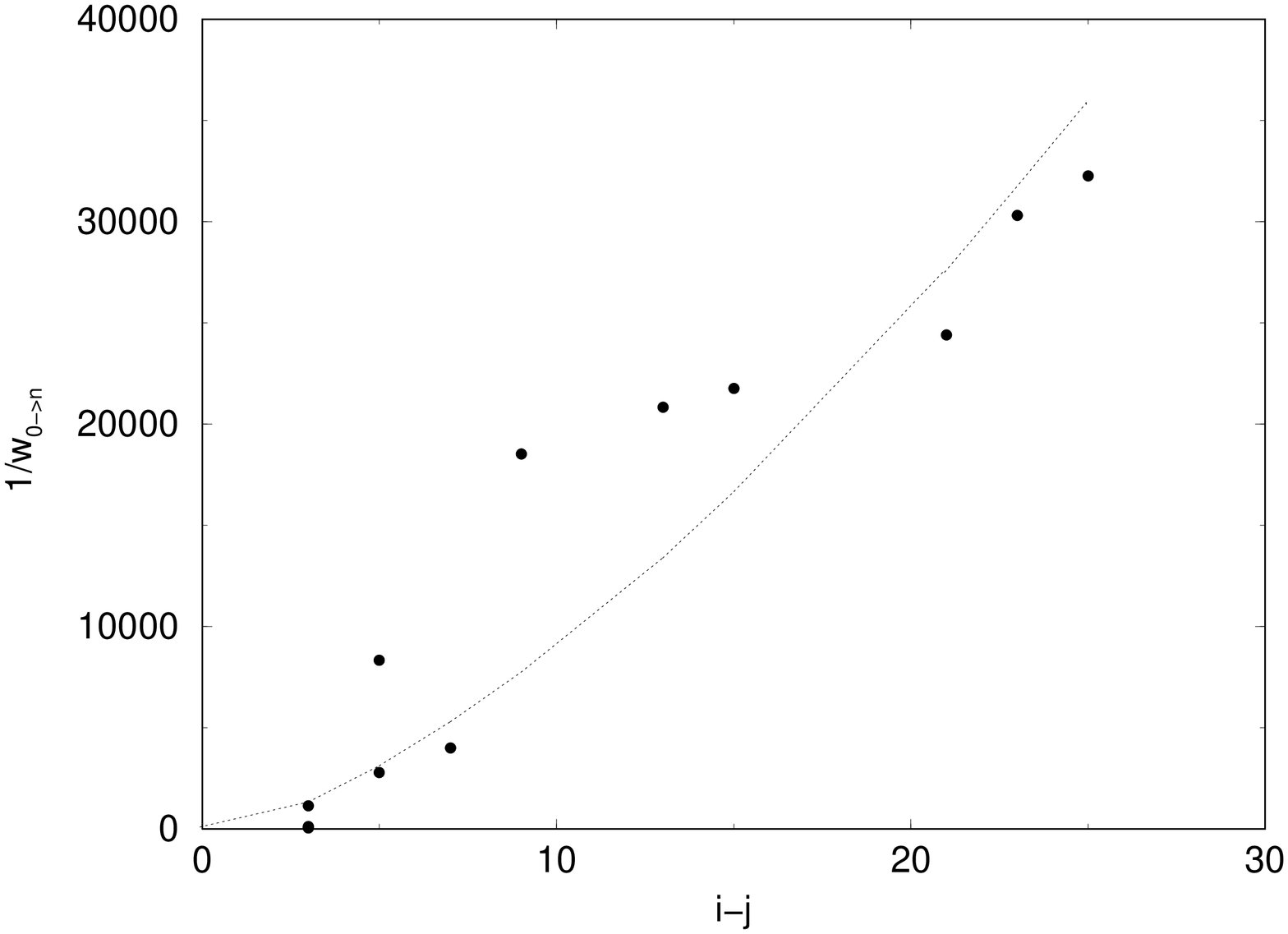,height=7cm,width=8.6cm,angle=-90}}
\caption{The inverse of the parameter $\wa$ is plotted with respect to
the distance $|i-j|$ of the monomers involved in the bond, measured
along the chain (circles), compared to the probability that the two
monomers come close during a random search (dotted line), normalized
accordingly.}  
\end{figure}


\begin{thebibliography}{99}
\bibitem{footnote1} One has to
remember that move attempts have not to violate the double occupancy
restriction, so it is not trivial that all monomers are attempted to move
with the same probability. In particular, after the chain has reached its native
state, this is no longer true, in that the movement of bulk sites is attempted
with a lower probability rather than surface sites.
\bibitem{footnote2} This seems not to be true if the interaction matrix is changed
to a new one whose elements are, in average, repulsive 
(cf. ref \protect\cite{sh_nucleus}).
\bibitem{footnote3} For longer loops
one can use the standard analytical
expressions available in the literature (cf. e.g. ref. \cite{degennes})
\end{thebibliography}
\end{document}